\documentstyle[preprint,aps]{revtex}

\begin{document}
\title{{\bf Linear to quadratic crossover of Cooper-pair dispersion relation}}
\author{Sadhan K. Adhikari$^1$\thanks{%
corresponding author $-$ e-mail: adhikari@ift.unesp.br, Fax: +55 11 3177
9080, address: Instituto de F\'{\i}sica Te\'{o}rica, Universidade Estadual
Paulista, Rua Pamplona 145, 01405-900 S\~{a}o Paulo, SP, Brazil}, M. Casas$%
^2 $, A. Puente$^2$, A. Rigo$^2$, M. Fortes$^3$, M. A. Sol\'{\i}s$^3$, M. de
Llano$^4$, Ariel A. Valladares$^4$, O. Rojo$^5$}
\address{$^1$Instituto de F\'{\i}sica Te\'{o}rica, Universidade Estadual
Paulista, 
01405-900 S\~{a}o Paulo, SP, Brazil}
\address{$^2$Departament de F\'{\i}sica, Universitat de les Illes
Balears,
07071 Palma de Mallorca, Spain}
\address{$^3$Instituto de F\'{\i}sica, Universidad Nacional Aut\'{o}noma
de
M\'{e}xico,\\
Apdo. Postal 20-364, 01000 M\'{e}xico, DF, Mexico}
\address{$^4$Instituto de Investigaciones en Materiales, Universidad
Nacional
Aut\'{o}noma de M\'{e}xico \\
Apdo. Postal 70-360, 04510 M\'{e}xico, DF, Mexico}
\address{$^5$PESTIC, Secretar\'{\i}a Acad\'{e}mica \& CINVESTAV, IPN,
04430 
M\'{e}xico DF, Mexico }
\maketitle

\begin{abstract}
Cooper pairing is studied in three dimensions to determine its binding
energy for all coupling using a general separable interfermion interaction.
Also considered are Cooper pairs (CPs) with nonzero center-of-mass momentum
(CMM). A coupling-independent {\it linear} term in the CMM dominates the
pair excitation energy in weak coupling and/or high fermion density, while
the more familiar quadratic term prevails only in the extreme low-density
(i.e., vacuum) limit for any nonzero coupling. The linear-to-quadratic
crossover of the CP dispersion relation is analyzed numerically, and is
expected to play a central role in a model of superconductivity (and
superfluidity) simultaneously accommodating a BCS condensate as well as a
Bose-Einstein condensate of CP bosons.

PACS \#: 74.20.Fg; 64.90+b; 05.30.Fk; 05.30.Jp

\end{abstract}





\newpage

\section{ Introduction}

The large-momentum divergence in the Cooper pair (CP) problem \cite{coop}
was originally eliminated by a momentum-space cutoff introduced in what is
now known as the Bardeen-Cooper-Schrieffer (BCS) model interaction which was
also successfully used by BCS in the study of conventional low-temperature
superconductors \cite{bcs}. On the other hand, the short coherence length of
some high-$T_{c}$ superconductors \cite{high} possibly implies a stronger
interfermion interaction. \ In the limit of very strong coupling one gets
well-isolated ``diatomic molecules'' or dimers of fermions, as opposed to
strongly overlapping CPs in the weak-coupling limit of BCS superconductivity 
\cite{mra,mrb,mrc}. In strong coupling these dimers can conceivably undergo
Bose-Einstein (BE) condensation (BEC). Although there is considerable
controversy over the precise pairing dynamics in so-called ``exotic'' \cite
{Uec}\ superconductors, tracking this many-body problem from strong to weak
coupling---known as the BCS-Bose crossover---has promoted the understanding
of various properties of exotic materials \cite
{mra,mrb,mrc,bcsboseb,bcsbosec,bcsbosed,bcsbosei,Hauss,adh1,adhia,1964,1965,nsr}%
.

Here we focus on how Cooper pairing itself evolves from weak to strong
coupling. We also study the excitation of CPs with nonzero center-of-mass
momentum (CMM), which should play an important role in a superconducting or
superfluid transition of the many-fermion system simultaneously exhibiting
both BCS and BE condensates as in the formulation by Friedberg and Lee \cite
{Blattb} and more generally by Tolmachev \cite{Blatta}.

The BCS model interaction simulates a phonon-mediated force which is
effective in explaining the isotopic effect by appropriate momentum-space
cutoffs placed symmetrically on either side of the Fermi surface \cite{sch64}%
. However, exotic materials do not exhibit such systematic isotopic effects.
Also, as coupling increases the chemical potential $\mu $ {decreases } in
value from the (positive) Fermi energy $E_{F}$ (its value at zero
temperature and zero interaction), and even turns negative as the so-called
Bose regime is entered \cite{mra,mrb,mrc}. The Fermi surface then washes
out, eliminating any physical motivation for the BCS model interaction with
Fermi-surface dependent cutoffs. Alternatively, in the BCS-Bose crossover a
renormalization procedure \cite{adhika,adhikb} may be used to handle the
large-momentum divergences. This leads to a renormalized dynamical model
expressible in terms of physical observables of the system rather than {\it %
ad hoc} cutoffs.

Here we derive renormalized $t$-matrix and Cooper equations for a pair of
fermions which move in vacuum and in the Fermi sea, respectively. In three
dimensions (3D) Cooper binding is expressed in terms of the two-fermion
scattering length in vacuum. For a CP with a nonzero CMM we define a pair
excitation energy as the (positive) difference between the CP binding energy
at zero and at a finite CMM. For high fermion density and any coupling only
a {\it linear} term in CMM dominates \cite{fujita,physc}\ the CP excitation,
which was in fact mentioned as far back as 1964 (Ref. \cite{sch64}, p. 33).\
At any coupling and for vanishing fermion-number density a {\it quadratic}
term dominates which is just the kinetic energy of the composite pair. The
crossover from a linear to a quadratic dispersion for the pair excitation
energy is then illustrated via numerical calculations.

The CP dispersion relation enters the BE distribution function of the boson
number equation from which $T_{c}$ is extracted \cite{fw}. The linear CP
dispersion relation for weak coupling leads to novel phase-transition
properties in either a heuristic \cite{pla} or a first-principles \cite{pret}
BEC picture of superconductivity as BE-condensing CPs. \ It is common
knowledge that BEC is possible only for dimensions $d>2$ for bosons with
quadratic dispersion; this limitation reappears in virtually all BEC schemes
thus far applied to explain superconductivity \cite{Hauss,nsr,Blattb,Blatta}%
. But for bosons with a {\it linear} dispersion relation found here in weak
and medium coupling, BEC can now occur for all $d>1$. This should be
relevant in models of superconductivity encompassing both BCS and BE
condensates \cite{Blattb,Blatta}.

In Sec. II the two-body problem is formulated in vacuum for a short-range,
separable interaction. In Sec. III the renormalized CP equation is derived
for nonzero CMM. In Sec. IV the CP dispersion relation in CMM is obtained
numerically. Finally, Sec. V offers discussion and Sec. VI conclusions.

\section{Two-body problem in vacuum}

Consider $N$ fermions in a box with sides of length $L$ that interact via an
S-wave short-range, attractive (rank-one) separable potential in 3D of the
form \cite{nsr} 
\begin{equation}
V_{pq}=-(v_{0}/L^{3})g_{p}g_{q},  \label{1}
\end{equation}
where $v_{0}\geq 0$ is the interaction strength and $g_{p}$ are
dimensionless form factors $g_{_{p}}=(1+p^{2}/p_{0}^{2})^{-1/2}$ \cite{nsr},
where the parameter $p_{0}$ is the inverse range of the potential. \ Such an
interaction model may mimic a wide variety of short-range effective
interactions: a force mediated by phonons, plasmons, excitons, magnons, etc.
or even a purely electronic interaction. Here $p_{0}\rightarrow \infty $
implies $g_{_{p}}=1$ and corresponds to a zero-range potential. The
advantage of potential (\ref{1}) is that many problems then yield analytic
solutions. The BCS model interaction is a special case of (\ref{1}) when $%
g_{p}$ is constant in the interval $E_{F}-\hbar \omega _{D}<$ $\hbar
^{2}p^{2}/2m<E_{F}+\hbar \omega _{D}$ and zero otherwise, where $\omega _{D}$
is the Debye frequency. More realistic potentials can be approximated by a
rank-N separable potential \cite{AD}.

The Lippmann-Schwinger equation for the $t$-matrix with potential (\ref{1})
between two fermions each of mass $m$ in free space is 
\begin{equation}
t_{pq}(E)=V_{pq}+\sum_{{k}}V_{pk}\frac{1}{E-\hbar ^{2}k^{2}/m+i0}t_{kq}(E),
\label{2}
\end{equation}
where $E$ is the two-particle energy. \ For potential (\ref{1}) the solution
of Eq. (\ref{2}) is \cite{AD} $\vspace{-0.01in}$ 
\begin{equation}
\vspace{-0.03in}\vspace{-0.05in}\smallskip t_{pq}(E)=\frac{g_{p}g_{q}}{%
\vspace{-0.1in}\vspace{0in}-\mathstrut \mathop{\displaystyle {L^3 \over
v_{0}}}-\mathstrut \mathop{\displaystyle \sum }\limits_{{\ k}}\medskip
\medskip \smallskip {\displaystyle{\frac{g_{k}^{2}}{E-\hbar ^{2}k^{2}/m+i0}}}%
}.  \label{3}
\end{equation}
In the limit $L\rightarrow \infty $, the momentum sum may be replaced by an
integral 
\begin{equation}
\sum_{{\ k}}\rightarrow \nu \frac{L^{3}}{(2\pi )^{3}}\int d^{3}k,
\label{sum}
\end{equation}
where $\nu $ is the spin degeneracy. Using (\ref{sum}), the zero-energy,
on-shell $t$-matrix is given in terms of the (S-wave) scattering length $a$
by \cite{AD} 
\begin{equation}
t_{00}(0)=\frac{4\pi \hbar ^{2}a}{m\nu L^{3}},  \label{sc}
\end{equation}
Eq. (\ref{3}) for $E=0$ then becomes 
\begin{equation}
\frac{\nu }{4\pi }\frac{mL^{3}}{\hbar ^{2}a}=-\frac{L^{3}}{v_{0}}+\sum_{{\ k}%
}\frac{g_{k}^{2}}{\hbar ^{2}k^{2}/m},\;\quad \;\;\text{ }  \label{5}
\end{equation}
since $g_{0}\equiv 1$, where the $i0$ term in the denominator is unnecessary
as the sum no longer diverges in the small $k$ limit.

\section{Two-body (Cooper) problem in Fermi sea}

The CP equation for two fermions above the Fermi sea with momentum
wavevectors ${\bf k}_{1}$ and ${\bf k}_{2}$ is given by 
\begin{equation}
\biggr[{\frac{\hbar ^{2}k^{2}}{m}-E_{K}+\frac{\hbar ^{2}K^{2}}{4m}}\biggr]%
C_{k}=-\sum_{{\ q}}{}^{^{\prime }}{}V_{kq}C_{q},  \label{6}
\end{equation}
where ${\bf k}\equiv \frac{1}{2}({\bf k}_{1}-{\bf k}_{2})$ is the relative,
and ${\bf K}\equiv {\bf k}_{1}+{\bf k}_{2}$ the center-of-mass, momentum
wavevectors, $E_{K}\equiv 2E_{F}-\Delta _{K}$ the total pair energy, $\Delta
_{K}\geq 0$ the CP binding energy, $C_{q}$ its momentum-space wave function,
and the prime on the sum implies restriction to states {\it above} the Fermi
surface: viz., $|{\bf k}\pm {\bf K}/2|>k_{F}$, where $k_{F}$ is the Fermi
wave number. For potential (\ref{1}), Eq. (\ref{6}) can be solved and $%
\Delta _{K}$ determined from 
\begin{equation}
\sum_{{\ k}}{}^{^{\prime }}{}\frac{g_{k}^{2}}{\hbar ^{2}k^{2}/m+\Delta
_{K}-2E_{F}+\hbar ^{2}K^{2}/4m}=\frac{L^{3}}{v_{0}}.  \label{7}
\end{equation}
Although the summand in Eq. (\ref{7}) is angle-independent, the restriction
on the sum arising from the full Fermi sea is a function of the relative
wave vector ${\bf k}$, and therefore angle-{\it dependent}. The interaction
strength $v_{0}$ in Eq. (\ref{7}) may then be eliminated by combining with
Eq. (\ref{5}). Thus, in terms of the scattering length $a$, the {\it %
renormalized CP equation} is 
\begin{equation}
\sum_{{\ k}}{}^{^{\prime }}\frac{g_{k}^{2}}{\hbar ^{2}k^{2}/m+\Delta
_{K}-2E_{F}+\hbar ^{2}K^{2}/4m}-\sum_{{\ k}}\frac{g_{k}^{2}}{\hbar
^{2}k^{2}/m}=-\frac{m\nu L^{3}}{4\pi \hbar ^{2}}\frac{1}{a}.  \label{9}
\end{equation}
For an attractive interaction with $g_{k}=1$ the sums in Eq. (\ref{9}) have
large-momentum divergences. However, each of the sums of Eq. (\ref{9})
diverges in the same fashion so that the difference is finite. A similar
renormalization of a general scattering equation can be performed \cite
{adhika,adhikb}. Also, renormalized BCS gap and number equations have been
used in the BCS-Bose crossover problem \cite
{mra,mrb,mrc,bcsboseb,bcsbosec,bcsbosed,bcsbosei,Hauss,adh1,adhia,1964,1965,nsr}%
. We shall determine the binding energy $\Delta _{K}$ with Eq. (\ref{9})
employing $a$ as interaction coupling parameter instead of the potential
parameter $v_{0}$. In what follows variables are dimensionless and expressed
in terms of $k_{F}$ or $E_{F}\equiv \hbar ^{2}k_{F}^{2}/2m$, viz., $\xi
\equiv k/k_{F}$, $\tilde{K}\equiv K/k_{F}$, $\tilde{\Delta}_{K}\equiv \Delta
_{K}/E_{F}$, $\tilde{a}\equiv ak_{F}$, etc.\

The sums in Eq. (\ref{9}) are transformed into integrals using Eq. (\ref{sum}%
). The restriction in the first term of Eq. (\ref{9})\ arising from the full
Fermi sea leads to two expressions depending on whether $\tilde{K}\equiv
K/k_{F}$ is $<2$ or $>2$, as discussed in the Appendix, 
\begin{equation}
\int_{0}^{\pi /2}d\theta \sin \theta \biggr[\int_{0}^{\Lambda }d\xi g_{\xi
}^{2}-\int_{\xi _{0}(\theta )}^{\Lambda }d\xi \frac{\xi ^{2}g_{\xi }^{2}}{%
\xi ^{2}-\alpha _{K}^{2}}\biggr]=\frac{\pi }{2\widetilde{a}},\text{ \ \ \ \
\  }\widetilde{K}<2,  \label{11}
\end{equation}
\begin{eqnarray}
&\int_{0}^{\pi /2}&d\theta \sin \theta \left[ \int_{0}^{\Lambda }d\xi g_{\xi
}^{2}-\int_{0}^{\Lambda }d\xi \frac{\xi ^{2}g_{\xi }^{2}}{\xi ^{2}+\beta
_{K}^{2}}\right]   \nonumber \\
&+&\int_{0}^{\theta _{0}}\sin \theta d\theta \int_{\xi _{0}^{\prime }(\theta
)}^{\xi _{0}(\theta )}d\xi \frac{\xi ^{2}g_{\xi }^{2}}{\xi ^{2}+\beta
_{K}^{2}}=\frac{\pi }{2\widetilde{a}},\text{ \ \ \ \ \ \ \ \ }\widetilde{K}%
>2,  \label{11bis}
\end{eqnarray}
where $\alpha _{K}^{2}\equiv 1-\tilde{\Delta}_{K}/2-\tilde{K}^{2}/4$, $\beta
_{K}^{2}=-\alpha _{K}^{2}$, $\xi _{0}(\theta )\equiv \sqrt{1-\tilde{K}%
^{2}\sin ^{2}\theta /4}+\tilde{K}\cos \theta /2$, $\xi _{0}^{\prime }(\theta
)\equiv -\sqrt{1-\tilde{K}^{2}\sin ^{2}\theta /4}+\tilde{K}\cos \theta /2$, $%
\theta _{0}=\arcsin (2/\tilde{K})<\pi /2$, with $\theta $ the angle between $%
{\bf k}$ and ${\bf K}$. To deal with the large-momentum divergences we have
introduced a finite upper limit $\Lambda $ and eventually let $\Lambda
\rightarrow \infty $. The momentum-space integrals in Eqs. (\ref{11}) and (%
\ref{11bis}) are easily performed for a zero-range interaction, i.e., with $%
g_{\xi }=1$. For CPs with $\widetilde{K}=0$, only the first equation is of
concern. Since in this case $\xi _{0}(\theta )=1$ and $\alpha _{K}^{2}\equiv
\alpha _{0}^{2}\equiv -\beta _{0}^{2}=1-\widetilde{\Delta }_{0}/2$ \medskip
, one arrives at 
\begin{equation}
1+\frac{\alpha _{0}}{2}\ln \left| \frac{1-\alpha _{0}}{1+\alpha _{0}}\right|
=\frac{\pi }{2\widetilde{a}},\;\;\;\;\;\;\alpha _{0}^{2}>0,  \label{3a}
\end{equation}
\begin{equation}
\frac{\pi }{2}\beta _{0}+1-\beta _{0}\arctan \frac{1}{\beta _{0}}=\frac{\pi 
}{2\widetilde{a}},\;\;\;\;\quad \alpha _{0}^{2}<0.  \label{3b}
\end{equation}
These two equations relate the dimensionless scattering length $-\infty <%
\widetilde{a}\equiv k_{F}a<+\infty $ to the dimensionless CP binding energy
for zero CMM $\widetilde{\Delta }_{0}\equiv \Delta _{0}/E_{F}$ for all
coupling. Transcendental equation (\ref{3a}) can be solved for CP binding
for weak coupling. In this limit the argument of the logarithm in Eq. (\ref
{3a}) reduces to $\widetilde{\Delta }_{0}/8,$ and one obtains 
\begin{equation}
\widetilde{\Delta }_{0}\rightarrow (8/e^{2})\exp (-\pi /|\widetilde{a}%
|)\;\;\;\text{\ (weak\thinspace \thinspace coupling),}  \label{19b}
\end{equation}
a limit first reported by Van Hove\cite{vanhove}. One can also find $%
\widetilde{\Delta }_{0}$ for strong coupling or for $\widetilde{\Delta }%
_{0}\rightarrow \infty .$ Since in this limit $\beta _{0}\arctan (1/\beta
_{0})\approx 1$, we obtain from Eq. (\ref{3b}) 
\begin{equation}
\widetilde{\Delta }_{0}\rightarrow 2/\widetilde{a}^{2}\;\;\;\;\;\text{
(strong\thinspace \thinspace coupling). }  \label{19c}
\end{equation}
Equations (\ref{3a}) and (\ref{3b}) were solved numerically to obtain the
exact functional dependence of $\widetilde{\Delta }_{0}$ on $1/\widetilde{a}$
, and this is compared with asymptotic forms Eqs. (\ref{19b}) and (\ref{19c}%
). In Fig. 1 we plot $\widetilde{\Delta }_{K}$ {\it vs }$1/\widetilde{a}$
spanning weak to strong coupling. One sees how the asymptotic form given by
Eq. (\ref{19b}) (short-dashed line) coincides with the exact $\widetilde{K}=0
$ result in weak coupling, whereas Eq. (\ref{19c}) (long-dashed curve) is
also quite accurate over a sizeable region for strong coupling.

\section{ Cooper pair dispersion curves}

Equations (\ref{11}) and (\ref{11bis}) are valid for all $K\geq 0$ and all
coupling. They can be solved numerically for CP binding $\Delta _{K}$ for
any $K$. Before discussing numerical results we derive analytically the
small-CMM behavior for zero range using $g_{\xi }=1$ for weak coupling in
Eq. (\ref{11}) which we take both for a small but non-zero $\widetilde{K}$
and for $\widetilde{K}=0$, and then subtract one equation from the other. A
small-CMM expansion of the resultant equation leads to the weak coupling
expression 
\begin{equation}
{\lim\limits_{\widetilde{\Delta }_{0}\ll 1}}\varepsilon {_{K}}=\frac{1}{2}%
\hbar v_{F}K+O(K^{2})+...,  \label{dk3}
\end{equation}
where a positive {\it CP excitation energy } $\varepsilon _{K}\equiv {%
(\Delta _{0}-\Delta _{K})}$ has been defined, and the Fermi velocity $v_{F}$
is given by $E_{F}/k_{F}=\hbar v_{F}/2$. The coefficient of the linear term
depends {\it only} on properties of the Fermi sea and not on any parameters
of the potential. In contrast, the complete excitation energy does depend on
the coupling parameter $\tilde{a}\equiv k_{F}a.$

It is this excitation energy that must enter the BE distribution function in
determining the critical temperature in a picture of superconductivity as a
BEC of CPs \cite{nsr,Blattb,Blatta,pla}. The leading term in (\ref{dk3}) is
linear in the CMM, followed by a quadratic term. But it is only for
sufficiently small fermion density, i.e., when $k_{F}$ or $E_{F}\rightarrow 0
$, and for {\it any }nonzero coupling, that the quadratic term dominates,
viz., 
\begin{equation}
\varepsilon {_{K}}\rightarrow \frac{\hbar ^{2}K^{2}}{2(2m)}\text{, }
\label{scl}
\end{equation}
the familiar nonrelativistic kinetic energy of the composite pair of mass $2m
$ and CMM h\hskip-.2em\llap{\protect\rule[1.1ex]{.325em}{.1ex}}\hskip.2em $K$%
. As mentioned, it is this dispersion relation that has been assumed in
virtually all BEC studies of superconductivity \cite{Hauss,nsr,Blattb,Blatta}%
. However, recent calculations of root-mean-square radii in two (2D) and
three (3D) dimensions in the BCS-Bose crossover scheme, when compared with
experimental coherence lengths of several typical 2D cuprates \cite{1964} as
well as of 3D materials \cite{1965}, suggest that they are describable well
within the BCS (weak-coupling) regime and away from the Bose
(strong-coupling) one. This implies that the {\it linear} approximation to
the dispersion relation would be relevant in these cases, and that perhaps a
more general description of the BEC of CPs for all coupling might require
the exact dispersion relation.

In Fig. 2 we display the reduced CP excitation energy $\varepsilon {%
_{K}/\Delta }_{0}$ as a function of reduced CMM $K/k_{F}$ for zero- and
finite-range potentials. Note that the CPs {\it break up} when $\varepsilon {%
_{K}/\Delta }_{0}\equiv (\Delta _{0}-\Delta _{K})/\Delta _{0}=1$, i.e., when 
$\Delta _{K}$ vanishes and turns negative. These points are marked by dots
in Fig. 2. For zero range we solve Eqs. (\ref{11}) and (\ref{11bis}) for $%
g_{\xi }=1$ and for typical values of $\Delta _{0}/E_{F}$ spanning weak to
strong coupling. For finite range we display results using $%
g_{p}=(1+p^{2}/p_{0}^{2})^{-1/2}$ with $p_{0}=k_{F}$ (i.e., range $1/p_{0}$
of the order of the average interfermion spacing $\sim 1/k_{F}$). \ Also
shown in Fig. 2 is the quadratic approximation in $K$ as given by Eq. (\ref
{scl}). \ We have labeled the curves by $\Delta _{0}/E_{F}$ as we found that
the zero-range curves are closer to the corresponding finite-range ones than
if they are labeled by $1/k_{F}a$ as in Fig. 1. The linear approximation Eq.
(\ref{dk3}) is valid only in the very weak-coupling and/or high fermion
density limit. The quadratic term dominates only at vanishing density, for
any nonzero coupling. For finite-range interaction the crossover in Fig. 2
is characterized by an inflection point with positive slope while for
zero-range there is no such inflection point.

\section{Discussion}

\bigskip The single-CP problem treated here may appear academic at first.
However, it has serious consequences. Our CPs are taken as ``bosonic'' even
though they do {\it not} obey (Ref.\cite{sch64} p. 38) Bose commutation
relations. This is because for a given $K$ they have {\it indefinite}
occupation number since for fixed $K$ there are (in the thermodynamic limit)
an indefinitely large number of allowed (relative wavenumber) $k$ values. \
Hence, for any coupling---and thus any degree of overlap between them---CPs
do in fact obey the Bose-Einstein distribution from which BEC is determined.
There have been attempts \cite{Blattb,Blatta} to formulate the superfluid
and superconducting transition problem in a many-fermion system by
accommodating both BE and BCS condensed phases. In these studies, the BCS
(BE) condensed phase dominates for weak (strong) coupling. For intermediate
coupling one could have both types of condensation with a certain density of
CP bosons (fermions) available for BE (BCS) condensation. However, the full
boson dispersion relation should be used to calculate the BEC transition
temperature.

The linear dispersion relation of a CP {\it should not }be confused with the
linear dispersion of Anderson-Bogoliubov-Higgs (ABH) many-body excitation
phonon-like modes. Collective modes in a superconductor were studied since
the late 1950's by several workers. \ A more recent treatment for 1D, 2D and
3D is available \cite{Bel} which confirms the linear ABH form $\hbar v_{F}K/ 
\sqrt{d}$ for $d$ =1, 2 or 3 in the zero-coupling limit. ABH phonons (like
photons or plasmons, etc.) {\it cannot }suffer a BEC as their number is
always indefinite. The number of CPs, on the other hand, is {\it fixed},
say, at half the number of (pairable) fermions if all of these are imagined
paired.

The above crossover of linear to quadratic forms of the CP dispersion
relation was also found in 2D \cite{physc,mex} so that we include both 2D
and 3D cases in the following discussion of the BEC transition temperature $%
T_{c}$. The 2D case is specially interesting as $T_{c}$ is zero for the
usual quadratic dispersion relation of CP bosons but {\it nonzero} for
linear dispersion.

The general BEC $T_{c}$ formula for bosons in any dimension $d$ and with a
general boson dispersion relation $\varepsilon _{K}=C_{s}K^{s}$, for $s$\ $>0
$, is \ given \cite{pla} for $d>s$ by
\begin{equation}
T_{c}=\frac{C_{s}}{k_{B}}\left[ \frac{s\,\Gamma (d/2)\,(2\pi )^{d}n_{B}}{%
2\pi ^{d/2}\,\Gamma (d/s)\zeta ({d/s})}\right] ^{s/d},  \label{gentc}
\end{equation}
but {\it vanishes} for $d\leq s$.  Here $n_{B}$ is the number density of
bosons of mass $m_{B}$, $k_{B}$ the Boltzmann constant, and $\zeta ({d/s})$
the Riemann Zeta function. \ For quadratic dispersion $s=2$, $C_{s}=\hbar
^{2}/2m_{B}$ and in 3D $\zeta (3/2)\simeq 2.612$ (\ref{gentc}) leads to the
familiar $T_{c}$ formula $T_{c}\simeq {\ 3.31\hbar ^{2}n_{B}^{2/3}/}%
m_{B}k_{B}$ \cite{fw} and to the fact that $T_{c}=0$ for all $d\leq 2$. For
the {\it linear} dispersion case $s=1$, 
and consequently $T_{c}=0$ for all $d\leq 1$ and $T_{c}>0$ for all $\ d>1$.
\ The latter is precisely the range of effective dimensionalities for all
known superconductors if one includes the quasi-1D organo-metallic Bechgaard
salts \cite{quai1d1,quai1d2}.

Before discussing the consequences of the $T_{c}$ formula (\ref{gentc}) with 
$s=1$ in superconductivity, we stress its limitations. Firstly, this $T_{c}$
formula with $s=1$ is derived with the linear dispersion relation
predominant for weak to moderate coupling, while the full correct dispersion
relation should be used in general. Secondly, in deriving this formula we
have taken the full momentum space of CP bosons so that the momentum
integrals run from 0 to $\infty $, whereas we have seen that the CPs break
up above some specific momentum value so the integrals should run only from
0 to the breakup $K_{0}$. Thirdly, the effect of unpaired fermions in the
background is ignored.\ Nevertheless, preliminary study shows \cite{pla}\
that once we remove these three limitations the result (\ref{gentc}) for $s=1
$ does not change drastically.


If one assumes that {\it all }fermions are paired into CP bosons so that the
boson density $n_{B}$ is $n/2$ with $n$ $\equiv N/L^{3}$ the fermion density
in the normal state, Eq. (\ref{gentc}) with $s=1$ leads to huge values of $%
T_{c}$ $\sim 10^{3}$ K for weak to intermediate coupling---the region of
interest in superconductivity even though $T_{c}$ empirically is at most
about $\sim $ $100$ K. However, the number of paired fermions vulnerable to
BEC is strongly coupling-dependent and generally \cite{pret}\ is only a
small fraction of all the fermions so that $n_{B}$ in Eq. (\ref{gentc}) is
effectively much smaller than $n/2$. \ Thus, a realistic $T_{c}$ is
certainly feasible. In the extreme weak coupling limit $n_{B}\rightarrow 0$,
driving the BEC $T_{c}$ to zero and allowing the BCS theory to be recovered
from analyses as in Refs. \cite{Blattb,Blatta}. For higher coupling $n_{B}$
increases so that one accommodates both BE and BCS condensates. \ One can
surmise that in a realistic theory of superconductivity BE and BCS
condensates play their respective roles. Elaborate calculations must still
be performed with the exact CP dispersion in order to find a more accurate $%
T_{c}$ for this many-body system.

\section{Conclusions}

The single CP problem with non-zero CMM is studied as it evolves (or {\it %
crosses} {\it over}) by varying the interfermion short-range pair
interaction from weak to strong. The CP excitation energy is exhibited as a
function of its CMM. For weak coupling the correct excitation energy is a 
{\it linear } dispersion relation in the CMM, which changes gradually to a 
{\it quadratic }relation as coupling increases and/or density is reduced to
the vacuum limit. These results will play a critical role in a model of
superconductivity that includes both BCS and BE condensates \cite
{Blattb,Blatta}. With a quadratic dispersion the BEC $T_{c}=0$ in 2D, from
which one might infer that BEC is irrelevant for quasi-2D cuprate
superconductivity. However, even in 2D, nonzero BEC transition temperatures
emerge for weak and medium coupling where the linear dispersion relation is
found to dominate \cite{mex,physc}, thus vindicating the relevance of BEC
for such materials. The pioneering attempts \cite{Blattb,Blatta} developing
a model of superconductivity accommodating BE and BCS condensates both
assumed the quadratic dispersion relation for {\it unbreakable} CPs for all
coupling. It would be interesting to reformulate those studies using the
proper dispersion relation with a {\it finite} breakup momentum. Lastly,
although our study is based on a separable potential we expect our
conclusions on the CP dispersion to be valid more generally, and applicable
to superconductivity, and to neutral-atom superfluidity such as in liquid $%
^{3}$He and in trapped Fermi gases\cite{trapped}.

\section{Appendix}

The restriction that both fermions lie above the Fermi sea in Eq. (\ref{9})
can be written as 
\begin{equation}
\left( {\bf k}/k_{F}\pm {\bf K}/2k_{F}\right) ^{2}-1=\xi ^{2}\pm \xi 
\widetilde{K}\cos \theta +\widetilde{K}^{2}/4-1\geq 0,  \eqnum{A. 1}
\label{a1}
\end{equation}
where $\xi \;{\bf \equiv \;}k/k_{F}$ and $\widetilde{K}\equiv K/k_{F}$. The
equality leads to two pairs of roots in $\xi $, say $\xi _{1,2}=-a\pm b$ and 
$\xi _{3,4}=a\pm b$, where $a\equiv (\widetilde{K}/2)\cos \theta $, $b\equiv 
\sqrt{1-(\widetilde{K}^{2}/4)\sin ^{2}\theta }$, and $\theta $ is the angle
between ${\bf k}$ and ${\bf K}$.

For $\widetilde{K}<2$, $b>a$, one root of the two pairs is positive and the
other negative. Thus, Eq. (\ref{a1}) can be satisfied provided that $\xi $ $%
>\xi _{1},\xi _{2},\xi _{3},\xi _{4}$, or specifically, if $\xi >\xi
_{0}(\theta )\equiv a+b.$ For $\widetilde{K}>2$ and $\theta >\theta
_{0}\equiv \arcsin (2/\widetilde{K})$, $b$ becomes imaginary and Eq. (\ref
{a1}) is satisfied for all $\xi $. Therefore, there is no restriction in the
integration over $\xi $. However, for $\widetilde{K}>2$ and $\theta <\theta
_{0}$, $b<a$ the pair of roots $\xi _{1,2}$ are both negative while the pair 
$\xi _{3,4}$ are both positive (with $\xi _{3}>\xi _{4}$). Consequently, in
both cases Eq. (\ref{a1}) is satisfied only if $\xi $ is in the interval $%
[0,\xi _{0}^{\prime }(\theta )\equiv a-b],$ and in the interval $\left[ \xi
_{0}(\theta ),\infty \right] $ , respectively. Equation (\ref{9}) is
evaluated using these restrictions on the $\xi $ integration.

\smallskip

\section{Acknowledgments:}

M.deLl. thanks S. Fujita for discussions, D.M. Eagles for reading the
manuscript, and V.V. Tolmachev for extensive correspondence. Partial support
from UNAM-DGAPA-PAPIIT (Mexico) \# IN102198, CONACyT (Mexico) \# 27828 E,
DGES (Spain) \# PB95-0492 and FAPESP\ (Brazil) is gratefully acknowledged.

\section{Figure captions}

1. Dimensionless CP binding energy $\Delta _{K}/E_{F}$ calculated for
zero-range interaction using Eqs. (\ref{11}) and (\ref{11bis}) for different
CMM momentum $K$ ($K/k_{F}=0,1,$ and 3) versus the inverse dimensionless
scattering length $1/k_{F\text{ }}a$. \ Short-dashed straight line (\ref{19b}%
) holds for weak coupling while long-dashed quadratic one (\ref{19c}) is
valid for strong coupling.

2. Reduced CP excitation energy $\varepsilon _{K}/\Delta _{0}$ versus $%
K/k_{F}$ for different reduced couplings measured in terms of $\Delta
_{0}/E_{F}$ calculated from Eqs. (\ref{11}) and (\ref{11bis}) for zero-range
(full curves) and finite-range potential with $p_{0}=k_{F}$ (long-dashed).
Short-dashed curves are the quadratic term of Eq. (\ref{scl}). Dots denote
CMM wavenumbers $K_{0}$ where the CPs break up.

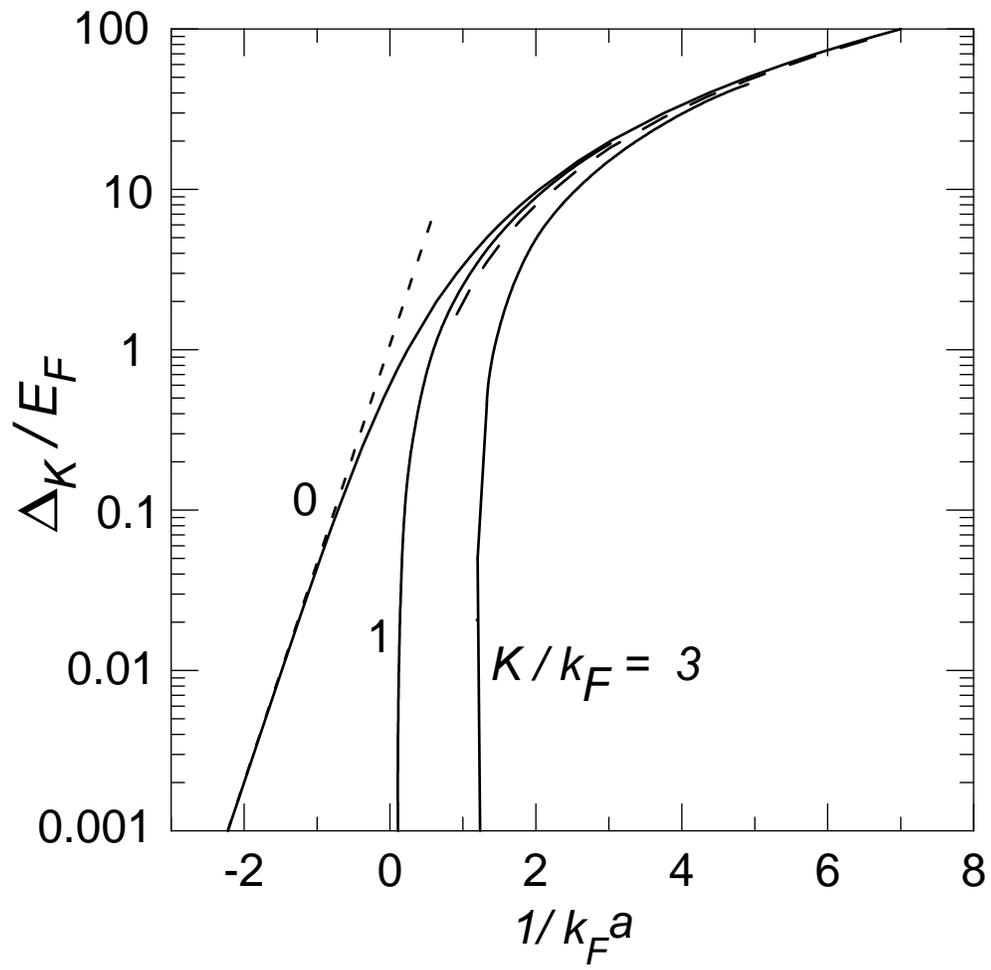
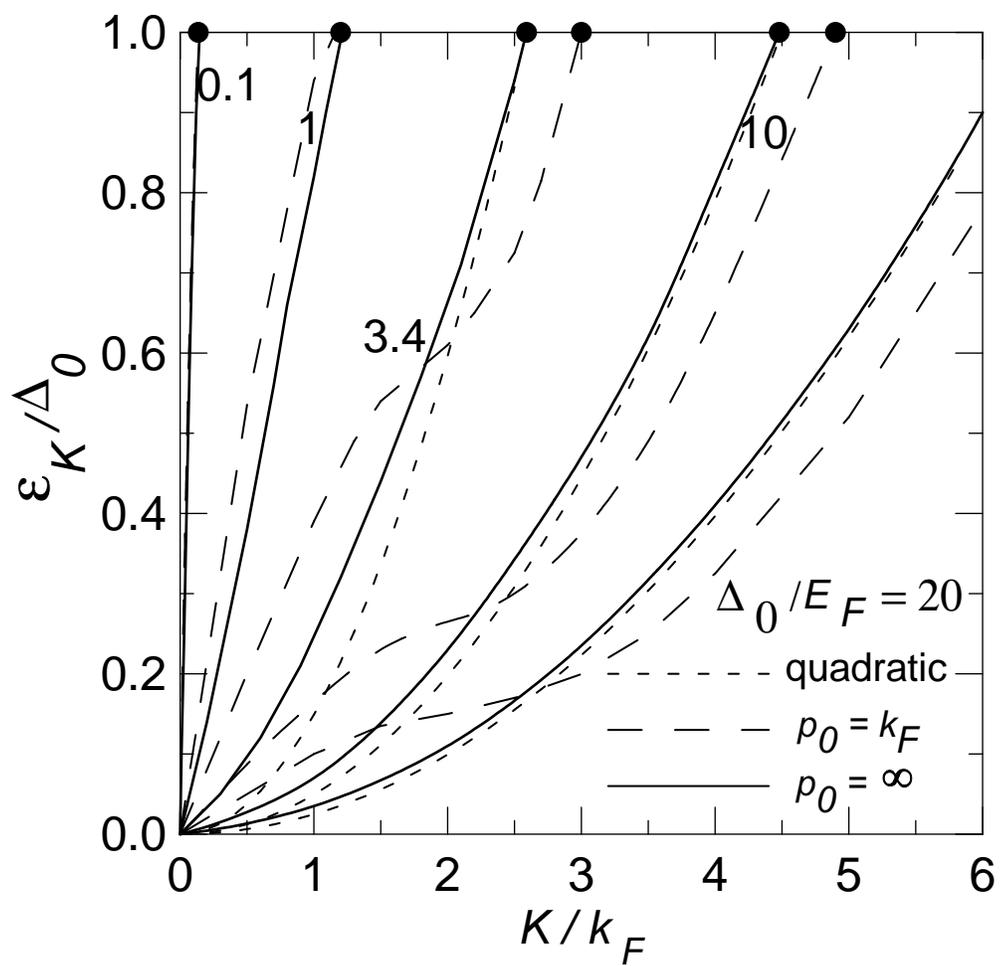
\end{document}